\documentclass[useAMS,usenatbib]{mn2e}
\usepackage{graphicx}
\usepackage{multicol}
\usepackage{amssymb}
\usepackage{times}
\usepackage{epsfig}
\usepackage{latexsym}
\usepackage{natbib}
\usepackage{footnote}
\usepackage[flushleft]{threeparttable}

\def\aj{AJ}                   
\def\araa{ARA\&A}             
\def\apj{ApJ}                 
\def\apjl{ApJ}                
\def\apjs{ApJS}               
\def\aap{A\&A}                
\def\aaps{A\&AS}              
\def\mnras{MNRAS}             
\def\prd{Phys.~Rev.~D}        
\def\nat{Nature}              

\newcommand{\zz}{$z_{850}$}
\newcommand{\ii}{$i_{775}$}
\newcommand{\vv}{$V_{606}$}

\title[LF of faint undetected $i$-dropout]{Constraining the luminosity
  function of faint undetected $i$-dropout galaxies$^{1}$}
\author[V. Calvi et al.]{V. Calvi$^{2,3}$\thanks{E-mail:
    valentina.calvi.1@studenti.unipd.it}, A. Pizzella$^{2,4}$,
  M. Stiavelli$^{3}$, L. Morelli$^{2,4}$, E. M. Corsini$^{2,4}$,
  E. Dalla Bont\`a$^{2,4}$, \and L. Bradley$^{3}$, and  A. M.
  Koekemoer$^{3}$\\ $^{1}$ Based on observations with the NASA/ESA
  Hubble space telescope, obtained at the Space Telescope Science
  Institute,\\ which is operated by the Association of Universities
  for Research in Astronomy, inc. under NASA contract
  NAS5-26555\\ $^{2}$ Dipartimento di Fisica e Astronomia
  ``G. Galilei'', Universit\`a di Padova, Vicolo dell'Osservatorio~3,
  I-35122 Padova, Italy.\\ $^{3}$ Space Telescope Science Institute,
  3700 San Martin Drive, Baltimore, MD 21218, United States.\\ $^{4}$
  INAF-Osservatorio Astronomico di Padova, Vicolo dell'Osservatorio~5,
  I-35122 Padova, Italy.}
\begin{document}

\date{Accepted ... Received ...}

\pagerange{\pageref{firstpage}--\pageref{lastpage}} \pubyear{2012}

\maketitle

\label{firstpage}

\begin{abstract}
\noindent

We present a new technique to quantify the light contribution coming
from the faint high redshift ($z\sim6$) galaxies below the detection
threshold of imaging data, set conventionally at $S/N=4.5$. We
illustrate the technique with an application to Hubble Space Telescope
Advanced Camera for Surveys images in the F775W and F850LP filters of
the Ultra Deep Field parallel field NICP12.  The aim of this analysis
is to extend by a few magnitudes the faint end of the luminosity
function at $z\sim6$. After masking all the detected sources in the
field we apply a Fast Fourier Transform to obtain the spatial power
spectrum of the background signal. The power spectrum permits us to
separate the background noise signal, the residuals due to the data
reduction of the wide field, and the overall signal produced by faint
galaxies. The ratio of the signal in the {\ii} and {\zz} bands is used
to estimate the contribution of the faint $i$-dropout objects. We rely
on extensive Monte Carlo simulations to characterize various sources
of uncertainty and quantify the number of faint $i$-dropout galaxies
in the field.  The analysis allows us to put constraints on the
luminosity function at $z\sim6$ down to {\zz}= 30 mag, 2.5 mag fainter
than with standard techniques on the same data. The data are
consistent with a faint end slope of the luminosity function of
$\alpha = -1.9$.  Assuming a specific set of values for the clumping
factor, escape fraction, and spectral energy distribution, we find
that the $z\sim6$ undetected galaxies down to $z_{850}=30$ mag could
have driven cosmic reionization.

\end{abstract}

\begin{keywords}
cosmology: early universe -- galaxies: high-redshift
\end{keywords}

\section{Introduction}

In the last few years we have begun to investigate the cosmic
reionization of hydrogen by using a variety of techniques.
Reionization was completed between $z \sim10$ and $z \sim6$
\citep{loebbarkana2001}, bracketed by the detection of Gunn-Peterson
troughs in the spectra of $z \sim 6$ quasars \citep{fan2006} and by
the Compton optical depth measured by the Wilkinson Microwave
Anisotropy Probe (WMAP; \citealt{jarosik2011}). However, the nature of
the sources responsible for cosmic reionization is still an open
question.
 
The UV photons expected to contribute to the reionization process come
from mainly two sources, star forming galaxies and quasars. The former
probably were fundamental for the hydrogen reionization, whereas the
latter likely dominated the helium reionization
\citep{shapirogiroux1987,madauhaardtrees1999}.  Accreting black holes
in the redshift range between $z = 6$ and $z = 8$ are rare and
possibly obscured by significant amounts of gas and dust, so their UV
emission probably does not contribute significantly to the
reionization process \citep{treister2011}.  According to
\cite{fan2001}, \cite{meiksin2005}, and \cite{schankarmathur2007}, $z
\sim 6$ quasars are not able to provide enough ionizing photons to
contribute significantly to reionization.

Population III stars are also unlikely to dominate the ionizing UV
production. In the standard scenario, in which only a single metal
free star per dark halo is formed, the contribution to reionization
given by Population III stars is only indirect. In fact, while the
cumulative number of ionizing photons produced by them is not enough,
they enrich the intergalactic medium of metals and allow the more
efficient formation of following generations of Population II stars
\citep{trentistiavelli2009}. These stars are likely located in faint
galaxies \citep{loebbarkana2001}. 

A large optical depth to electron scattering, consistent with the
Universe being reionized at $z\sim15$, was found by WMAP
\citep{bennett2003,komatsu2011}. Since there are still strong
uncertainties on several key parameters concerning reionization,
\cite{kuhlen2012} stated that it is not clear if galaxies could have
completed the reionization by $z\sim6$ and account for the large
optical depth to electron scattering found by WMAP-7. On the other
hand, most of the studies carried out in the last decade are focused
on the critical star formation rate density (SFRD).  The Lyman break
galaxies detected at $z \sim6$ \citep{stiavelli2004b} and $z \sim7$
\citep{bouwens2004,boltonhaenelt2007,oesch2009} in the Hubble Ultra
Deep Field (hereafter HUDF) are not bright enough to account for the
required ionizing budget.  However, more recently
\cite{finkelstein2012} have found that the detected galaxies at $z
\sim 6$ are sufficient to fully reionize the IGM if $C/f_{esc} \sim
10$.  According to \cite{bunker2010} and \cite{finkelstein2010}, the
observed $z \sim 7$ galaxies could have completed the reionization
process if $f_{esc} \ge 50\%$.  The observed faint-end slopes imply
large numbers of undetected faint galaxies which likely provide the
bulk of ionizing photons and played an important role in cosmic
reionization \citep{yanwindhorst2004,richard2008,
  mclure2010,trenti2010,bouwens2011b}.

Unfortunately, direct probing of the faint galaxy population is
outside present capabilities and will need to wait for the James Webb
Space Telescope (JWST). Thus, in order to make further progress in
this field before the launch of JWST, one needs to find ways to study
the faint end of the galaxy luminosity function at $z \ge 6$ using the
integrated light contribution of individually undetected
galaxies. This idea is not new and it was tried in the context of
high-$z$ studies using the Spitzer Infrared Array Camera (IRAC) and
Advanced Camera for Surveys (ACS) data \citep{kashlinsky2007} and is,
in fact, also analogous to the study of surface brightness
fluctuations of nearby galaxies \citep{tonryschneider1988}.

The aim of this paper is to describe a new technique to analyze the
contribution of faint, individually undetected, galaxies where we
indicate as undetected galaxies those galaxies characterized by being fainter than the typical signal-to-noise ($S/N$) ratio of 4.5 
\citep{oesch2009}.  We focus mostly on galaxies just below the
detection threshold as such galaxies are likely to exist, given that
the luminosity function (LF) of galaxies should not know the HUDF
magnitude limit, and they might be sufficient for reionization
\citep{yanwindhorst2004}.

These galaxies are too faint to be measured individually, but their
signal can be quantified using a Fourier Transform based technique.
The main difference here over the previous attempts at high redshift
is our special attention to systematic effects.  

 The Lyman break technique permits to identify star-forming galaxies
 at high redshift using color criteria derived from a multi-band
 photometry in the region across the Lyman-continuum discontinuity
 \citep{steidelhamilton1993,giavalisco1998}.  Comparing the flux of a
 galaxy within a set of two or three ad-hoc filters depending on the
 redshift, it can be noticed that no flux is collected by the bluer
 filter due to the Lyman $\alpha$ absorption, while the object is
 detected in the redder bands.

In this paper we focus on $i$-dropout galaxies at $z\sim 6$ that can
be studied using images obtained with the same camera and
characterized by, essentially, the same systematic effects.  Galaxies
at $z\sim6$ are expected to be undetected in the ACS F775W band
(hereafter \ii) image and detected in the ACS F850LP (hereafter
\zz) one, so they are named \ii-dropout galaxies.

The faint galaxy population we seek is detected by an excess in the
power spectrum in the {\zz} image with respect to that of the {\ii}
one, i.e., it is a Lyman break population.  

In Section 2, we describe the data used.  In Section 3, we explain the
data analysis and we present the main steps of the power spectrum
technique adopted to obtain the signal coming from the undetected
galaxies, which are bona fide $z \sim6$ candidates.  In Section 4, we
describe how to estimate the contribution to cosmic reionization using
Monte Carlo simulations, we discuss our results and their implications
for the reionization epoch.

\section{Data Set}
\label{The data}

\begin{table}
\begin{center}
\caption{Properties of the NICP12 data.}
\begin{threeparttable}
  \begin{tabular}{ccccc}
    \hline \hline 
Filter & {\it HST} & Exposure & Zero Point$^b$& $10\,\sigma$ Magnitude$^a$\\
& orbits$^a$&Time$^a$ [s]&[AB mag]&[AB mag]\\ 
\hline 
{\ii} & 23 & 54,000 & 25.654&28.44\\ 
{\zz} & 69 & 168,000 &24.862&28.47\\ 
\hline

\end{tabular}
\begin{tablenotes}
\item $^a$  From \cite{oesch2007}
\item $^b$ From \cite{demarchi2004}
\end{tablenotes}
\end{threeparttable}
\end{center}
NOTES: Col. 1: filter name. Col. 2: number of HST orbits. Col. 3:
total exposure time in seconds. Col. 4: zero point in AB
magnitude. Col 5: $10\,\sigma$ limiting magnitudes in AB magnitude
computed within apertures of 0.15\farcs radius.
 \label{tab:NICP12}
\end{table}

To date, the deepest optical data available are from the HUDF project
\citep{beckwith2006} and the Hubble Ultra Deep Field Follow-Up
(hereafter UDF05; \citealt{oesch2007}).  These ultra deep observations
of multiple fields have been obtained with ACS. In this first
application of our technique we selected the parallel field NICP12
rather than the main HUDF field. The reason for this is that the
NICP12 images were produced with a more advanced data reduction than
the one used in the main field \citep{oesch2007}. In particular, an
herringbone effect introduced inadvertently by the subtraction of a
combined, high $S/N$, bias frame (superbias) made from compressed bias
frames containing a small electronic noise component was eliminated as
well as electronic ghost images of the bright sources.  Thus, the
improved version of the NICP12 images (v2.0) is the cleanest
available. We are currently reprocessing the main HUDF field so as to
bring it to the same standard and when this field will be available we
plan to apply this analysis to it as well.

The coordinates of the center of the NICP12 field are R.A.=$03^{\rm
  h}33^{\rm m}03.60^{\rm s}$, Dec.=$-27^{\circ}41'01.80''$
(J2000.0). The characteristics of the observations for the {\ii} and
{\zz} images are specified in Table \ref{tab:NICP12}. Throughout this
paper we used the AB mag magnitude system.  The original images are
characterized by a pixel scale of $30\; {\rm mas}\;{\rm pixel}^{-1}$
and a side of 10500 pixels. They were rotated to have North up and East
left.

\section{Data Analysis}

A first step to reveal the contribution of undetected galaxies was to
mask all the light coming from stars and detected galaxies, including
their bright halos.  Once the appropriate mask had been created we
applied the power spectrum technique to analyze the different
contributions existing in the background.

\subsection{Creation of the Mask}
Since the MULTIDRIZZLE task \citep{koekemoer2002} works on dithered
observations of the same field and only a few pointings in the outer
regions of the field can be combined, we used the weight maps
associated to the v2.0 data to select the central region for the
following analysis.  The data we focused on were binned to have a
scale of 90 mas pixel$^{\rm{-1}}$ and a size of 3500 $\times$ 3500
pixels.  This rebinning is intended to limit the computational
volume, but it does not affect the final result, as discussed in
Section \ref{Sources of Contamination}.
 
First of all, we had to remove all the defects and objects that could
interfere with our analysis, such as the detected galaxies, their
bright halos, residual cosmic rays, and bad pixels.  Galaxies were
detected independently in the {\ii} and {\zz} images using the
SExtractor photometry package by \cite{bertinarnouts1996}, version
2.5.0. The SExtractor parameters were optimized to maximize the number
of detected galaxies while minimizing the number of spurious sources.
The detection threshold (DETECT\_MINAREA) was set to be a minimum of 5
connected pixels with an intensity of $0.55 \sigma$ (DETECT\_THRESH)
above the background \citep{oesch2007}.

\cite{oesch2009} considered only those galaxies detected with $S/N >
4.5$ to be reliable. Since we were interested in galaxies which are
too faint to be individually studied, but that produce a relevant
overall light contribution, we had to mask all the reliable single
galaxies and focus on the total contribution of the fainter ones. We
noted that, by doing this, we were sensitive to both galaxies just
below our detection threshold and much fainter ones. This was
intentional, as we were trying to constrain all contributions below
those of reliably detected galaxies.

Starting from the segmentation maps, one of the outputs of SExtractor,
we created two masks (one for each band) to reject all the detected
sources existing in the field of view.  Since most of the objects are
surrounded by a bright halo, we decided to enlarge the masked area to
avoid contaminating signal from bright galaxies. We reversed each mask
to be 1 where there are sources and 0 elsewhere, we convolved two
times the mask with a Gaussian filter with ${\rm FWHM}=30\; {\rm
  pixel} = 2 \farcs 7$ enlarging it each time by adding all the pixels
with a value $> 0.5$. Finally we reversed again the mask.  In
  this way we added $\sim18-19\%$ more pixels to the mask.

 We obtained the final mask by merging the masks in the two passbands
 and convolving again. The masked area corresponds to 16.8\% of the
 field.

\subsection{The Power Spectrum Technique}
A description of random fluctuations is fundamental to extract
information hidden in the background of images and to provide a
quantitative measure for the comparison with simulations.  The spatial
power spectrum is the key of our method, allowing us to determine the
amplitude of surface brightness fluctuations and to distinguish the
sources of noise from the fluctuations due to the unresolved galaxies.
The power spectrum we mention is not the same applied first by
\cite{harrison1970} and \cite{zeldovich1972}, and then by
\cite{white1994} to study the cosmic microwave background (CMB)
anisotropies. We are, instead, interested in the spatial distribution
of the faint $z\sim6$ galaxies, so as to be able to extend the faint
end of the luminosity function. For this reason we used the Fourier
transform to derive the spatial power spectrum of the luminosity
fluctuation due to these galaxies, similar to the work of
\cite{tonryschneider1988} and \cite{tonry1990}. Using this method we
are able to constrain the number of the faint galaxies (see, however,
comments at the end of Section \ref{Sec:different_contribution}).

A technique similar to ours was used for the first time by
\cite{tonryschneider1988}. They discovered that the distance of a
galaxy is inversely proportional to the amplitude of the luminosity
fluctuations due to unresolved red giant stars and they used the
spatial power spectrum to directly measure these fluctuations. The
main difference between their work and ours is the density of sources
per pixel. They were studying nearby galaxies that have a projected
density of stars of many tens per pixel so the fluctuations between
adjacent pixels have rms variations $\lesssim 10$\% of the mean
signal. On the contrary, we are dealing with images where the
projected density of faint galaxies is much less then one per pixel,
therefore we have smaller fluctuations and our results are more
sensitive to the presence of spurious signals.

Our approach consists in applying the IDL\footnote{Interactive Data
  Language is distributed by Exelis Visual Information Solutions.}  Fast
Fourier Transform (FFT) routine to the masked images to compute the
two-dimensional Fourier transform and, then, to derive the spatial
power spectrum of the signal.  Therefore, the power spectrum image
obtained is two-dimensional, but to plot it and to do all the
calculations in the following it is convenient to derive the radial
trend.

The power spectra for the two bands can be compared by calculating
their difference or ratio. The difference is the most natural choice,
but relies on an extremely accurate color calibration of the power
spectra in order to properly subtract the contribution of lower
redshift objects from the {\zz} power spectrum. The ratio allows
us to avoid the need of a very accurate and challenging color
calibration of the power spectra. Thus, from here on, we focus our
analysis on the ratio of power spectra.

\begin{figure}
\centering
\includegraphics[scale=0.45]{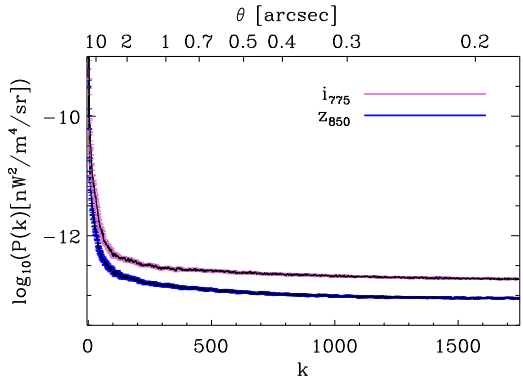}
\caption{Power spectra, with colored error bars, of the background
  signal of the {\ii} (top line with purple error bars) and {\zz}
  images (bottom line with blue error bars).}
\label{PS_iz}
\end{figure}

\begin{figure}
\includegraphics[scale=0.45]{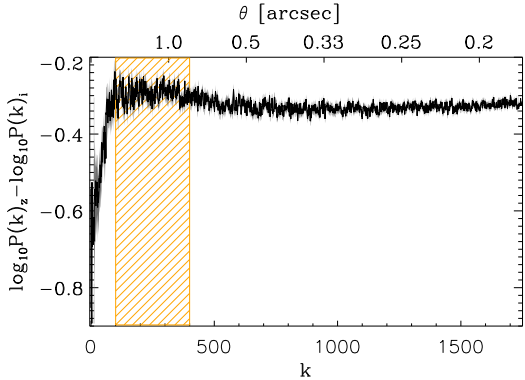}
\caption{Ratio between the {\zz} and {\ii} power spectra. The shaded
  grey area represents the errors associated to the ratio. The region
  between $k = 100$ and $k = 400$,  where the signal from the faint
  galaxies was  measured, is highlighted in orange.}
\label{PS_ratio}
\end{figure}

We obtained the power spectra for both the {\ii} and {\zz} images
(Figure \ref{PS_iz}). To highlight the light contribution coming from
the undetected galaxies, which are bona fide $z \sim 6$ candidates, we
examined the ratio between the {\zz} and {\ii} power spectra (Figure
\ref{PS_ratio}). These undetected galaxies are responsible for a peak
located between wavenumbers, k, of 100 and 400, shown in Figure
\ref{PS_ratio}. We tested the reliability of this feature using a
$\chi^2$ statistics. Comparing the values of the ratio between the
power spectra in the range of the peak, i.e. $100 \lesssim k \lesssim
400$, and in the range $1400 \lesssim k \lesssim 1700$ dominated only
by the noise, we obtained $\chi^2_{\nu} = 3.08$. This value, which can
be rejected with a confidence well beyond $10\sigma$, convinced us
that the peak is a real feature and it can not be ascribed to random
noise.

The angular scale $\theta$, plotted on the top x-axis of the figures, is
derived from $k$ as follows:
\begin{equation}
 \theta \; [{\rm arcsec}] =\frac{\rm D_{\rm frame}\; [\rm{pixel}]
  \cdot scale\; [\rm{arcsec} \; \rm{pixel^{-1}}]}{k}
\end{equation}

where $\rm D_{\rm frame}$ is the dimension of the frame in pixels and
scale is the pixel scale  of the image.

Following our technique, we can derive the number of faint galaxies
from the height of the signal excess they produce in the ratio between
the power spectra (Figure \ref{PS_ratio}) and by comparing it with the
one obtained from simulations (see Section \ref{sec:simulations}).

It must be noted that our result can be affected by a fraction of
interlopers at lower redshift. This is an issue affecting all studies
based on the dropout selection and it is due to the aliasing between
the Lyman break and the $\rm 4000$ \AA\/ break, as discussed by
\cite{dahlen2010}.  \cite{su2011}, on the basis of accurate
simulations, estimated that the fraction of $z\sim 1.2$ galaxies which
can be selected as $z\sim6$ population could be as high as 24\%. This
value is a pessimistic one since \cite{malhotra2005} found half that
number of contaminant galaxies, based on ACS grism observations.

\subsection{Different Contributions}
\label{Sec:different_contribution}
The use of the power spectrum technique requires a preparatory study
to understand how the different components of a scientific image
affect the final power spectrum.

The ratio between the spatial power spectra in Figure \ref{PS_ratio}
can be divided into three different parts, the first one at low wave
numbers ($k \lesssim 100$), the second one in the range of high wave
numbers ($k \gtrsim 400$) and the remaining one at intermediate wave
numbers, which corresponds to the contribution from the undetected
galaxies  at $z\sim6$.

The power spectrum of random noise is a white power spectrum
having approximately a constant value over all the range of wave
numbers. At high wave numbers, meaning at the pixel scale, the white
noise contribution is dominant. At low wave numbers the principal
contribution is due to residuals of the flat field correction process
(for more details on this see Section 3.4). The highlighted
region of the plot in Figure \ref{PS_ratio} is the most interesting
one,  where the signal is dominated by the contribution of the
  galaxies, therefore it permits us to infer the existence of
galaxies below  the usual detection limit.  In particular,
  the amplitude of the peak visible between $k = 100$ and $k = 400$ is
  proportional to the number of undetected galaxies in the field.

\subsection{Possible Sources of  Contamination}
\label{Sources of  Contamination}
There are many sources of uncertainties that can affect our results
and we will briefly list and characterize them.

\begin{figure}
\centering
\includegraphics[scale=0.45]{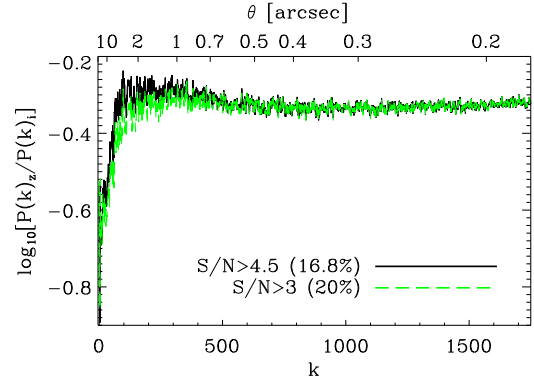}
\caption{Ratios between the {\zz} and {\ii} power spectra obtained
  masking all the sources detected by SExtractor with $S/N > 4.5$
  (black line) and $S/N > 3$ (green dashed line). The
  percentages indicate the masked area.}
\label{comparison_ratio_mask}
\end{figure}

{\it Mask:} The primary source of uncertainty in our results is the
size of the mask.  Our aim is to mask all the galaxies which can be
reliably and individually identified and to study the overall signal
coming from those which are too faint to be detected one by one.  If
we are masking to a signal level which is too low, we will not see any
signal coming from faint galaxies.  On the other hand, if the mask
brightness limit is too high the faint $z \sim 6$ undetected galaxies
will be overwhelmed by the light of the foreground galaxies and their
bright halos. The ratio between the power spectra obtained by masking
all the sources detected by SExtractor with $S/N > 4.5$ and $S/N >3$
are compared in Figure \ref{comparison_ratio_mask}. In the region
between $k=100$ and $k=400$ we notice that the peak due to the signal
of faint $i$-dropouts is significantly lower and almost invisible. As
the percentage of masked area is quite similar (16.8\% and 20\%), we
can infer that we are detecting a signal from galaxies barely below
our detection threshold rather than from a multitude of very faint
objects.

\begin{figure}
\centering
\includegraphics[scale=0.45]{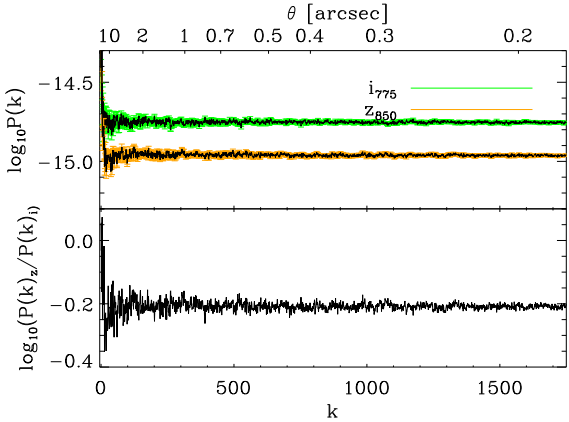}
\caption{Power spectra derived by combining the noise-only frames with
  MultiDrizzle (top panel) and their ratio (bottom panel). The
  flatness of the profile indicates that no features were introduced
  when combining the images with MultiDrizzle.}
\label{p12_noise_multidrizzle}
\end{figure}

\begin{figure}
\centering
\includegraphics[scale=0.45]{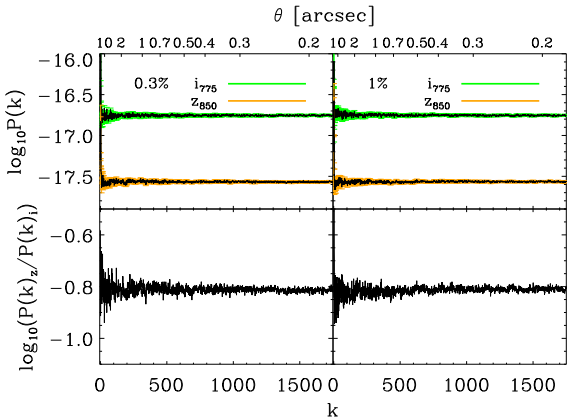}
\caption{ Power spectra derived by combining with MultiDrizzle
  noise-only frames after being multiplied by a random residual of
  flat field with a maximum of 3\% (top left panel) and 1\% (top right
  panel) and their ratio (bottom panel). As in Figure
  \ref{p12_noise_multidrizzle}, no spurious feature is introduced in
  the final ratio when combining the images with MultiDrizzle. }
\label{p12_noise_flat_multidrizzle}
\end{figure}

{\it Calibration Files and Dither Pattern:} The use of MultiDrizzle
could introduce spurious signals in the final scientific frame
obtained after the combination of all the {\it flt} files. To test
this, we downloaded all the {\it flt} data available for the NICP12
field in the \ii and \zz bands from the HST archive and replaced the
scientific frame with random noise. The resulting mock frames have no
sources and maintain the same coordinates of the real ones. Then, for
each filter we processed these noise-only frames with MultiDrizzle,
and we applied the power spectrum technique to the final scientific
images. As it can be noticed from Figure \ref{p12_noise_multidrizzle}
no features are introduced by MultiDrizzle neither in the single power
spectra (top panel), nor in the ratio (bottom panel). Moreover, we did
a similar test combining mock frames with noise multiplied by a random
residual of flat field (with a maximum of 0.3\% and 1\%) and, then,
applying the power spectrum technique.  It should be noted that
  the flat field residual in this case introduces a pixel-to-pixel
  uncertanty, not a large scale one. Once again, we concluded that
MultiDrizzle is not creating spurious signals (see Figure
\ref{p12_noise_flat_multidrizzle}) and, consequently, it is not
affecting the final result of our technique.

{\it Cosmic Ray Residuals:} The presence of cosmic ray residuals in
the frame results in a gentle increasing slope in the range of high
wave numbers, which depends on the intensity and number of pixels
affected by the cosmic rays.

{\it Zodiacal Light:} The zodiacal light is the dominant source of sky
brightness in the near-IR \citep{reach1997}. We were only interested
in the fluctuations due to the zodiacal light cloud changing in the
sky during the time of the observations. According to
\cite{kashlinsky2005}, considering two observations of the same field
taken six months apart and computing their difference, it is possible
to isolate the zodiacal light fluctuation from the contribution coming
from galactic and extragalactic objects. With this method
\cite{kashlinsky2005} were able to derive an upper limit for the
zodiacal light contribution and to affirm that it did not affect their
results on the cosmic infrared background fluctuations. Unfortunately,
a few issues prevented us from applying the same analysis to the HST
data.  First of all, observations taken six months apart imply
different orientations of the camera which could introduce
detector-related systematics leaving a signature in the power
spectrum. Moreover, ACS is affected by the contribution of scattered
sunlight whenever the angle is close to (or less than) $90^\circ$ and
this could determine a wrong estimate of the fluctuation ascribable to
zodiacal light.  Finally, at the moment we do not have observations of
our fields taken with such a long temporal delay, so we cannot study
relevant zodiacal fluctuations. On the basis of above, we decided to
assume a negligible zodiacal contribution following
\cite{kashlinsky2005} and to better handle the zodiacal light
fluctuation in our next paper, analyzing observations of the same
field obtained with a long temporal separation, so as to highlight
relevant differences in the zodiacal light.

{\it Flat Field Correction and Bias Subtraction Residuals:} If the
flat field correction or bias subtraction process were not correctly
done, the data could be affected by a diffuse signal not coming from
any real source. It can easily overcome the small amount of photons
produced by the undetected galaxies we are looking for. These
residuals modify the trend of each single power spectrum, increasing
its slope in the low and medium wave numbers ranges and making
difficult to see any peak in the power spectra ratio.

\begin{figure}
\centering
\includegraphics[scale=0.45]{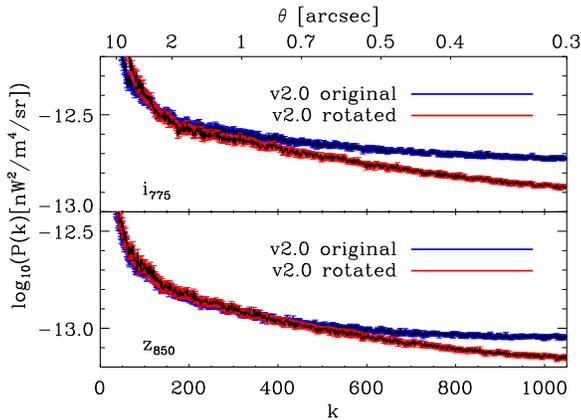}
\caption{Power spectra obtained from the \ii (top panel) and \zz
  (bottom panel) images with (bottom line with red error bars) and
  without (top line with blue error bars) applying a 45$^{\circ}$
  clockwise rotation. In each panel the power spectrum referring to
  the rotation was vertically shifted to match the other one at
  $k=125$ ($\rm D_{\rm galaxies}=2 \farcs 52$) to better show the
  effect of the rotation.}
\label{comparison_iz_rotation_paper}
\end{figure}

\begin{figure}
\centering
\includegraphics[scale=0.45]{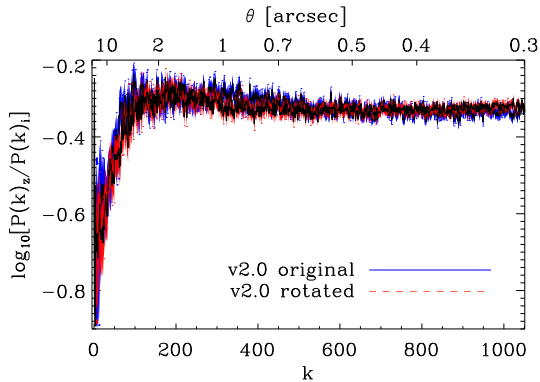}
\caption{Comparison of the ratios between the \zz and \ii power
  spectra for the v2.0 data with (red dashed error bars) and without
  (blue solid error bars) applying a 45$^{\circ}$ clockwise
  rotation. The rotation has no significant effect on the ratio.}
\label{comparison_ratio_rotation}
\end{figure}

{\it Rotation:} The images we used, oriented with North up and East
left, were obtained from MULTIDRIZZLE without introducing any
additional correlation between adjacent pixels \citep{casertano2000}
because of the use of the point kernel \citep{beckwith2006}. We did
not introduce any further rotation in our data, nevertheless, we
decided to investigate this effect for sake of completeness since this
is a technical paper describing all the aspects of our new
technique. If we were dealing with rotated frames it would be
important to consider the following effect.  The rotation of the
images before extracting the power spectrum determines a correlation
between adjacent pixels, which is detected as a clear slope at high
wave numbers of both the {\ii} and {\zz} power spectra (Figure
\ref{comparison_iz_rotation_paper}).  The magnified panel shows that
the power spectra of the original and rotated data present different
slopes, but, as shown in Figure \ref{comparison_ratio_rotation}, the
ratio of the power spectra is not significantly affected by rotation
since the slope introduced is exactly the same in both the {\ii} and
{\zz} power spectra. The ratio of the rotated data fits extremely well
with that of the unrotated ones.

\begin{figure}
\centering
\includegraphics[scale=0.45]{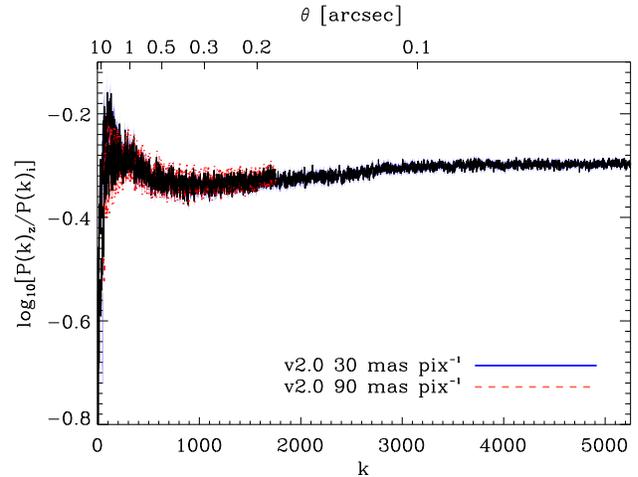}
\caption{Ratios between the {\zz} and {\ii} power spectra obtained for
  the v2.0 images with a pixel scale of $30$ (blue solid line error
  bars) and $90\; {\rm mas}\;{\rm pixel}^{-1}$ (red dashed line error
  bars). By binning the data the power spectrum covers a smaller wave
  number range, but the different pixel scale is not affecting the
  features we are interested in.}
\label{comparison_ratio_binning}
\end{figure}

{\it Binning:} To test the effect of data binning we compared the
results obtained from the images with a pixel scale of $90\; {\rm
  mas}\;{\rm pixel}^{-1}$ and a dimension of 3500 pixel each side, and
the original images, characterized by a pixel scale of $30\; {\rm
  mas}\;{\rm pixel}^{-1}$ and a 10500 pixel side.  By binning the data, we lost details
and so the power spectrum covers a more limited range of wave numbers.
However, the ratio between the power spectra plotted in Figure
\ref{comparison_ratio_binning} shows that the white power spectrum is
well reproduced by both the $30$ and the $90\; {\rm mas}\;{\rm
  pixel}^{-1}$ cases.

{\it Groups of Galaxies:} It is known that galaxies at high redshift
are not necessarily randomly distributed, but can be
grouped. \cite{overzier2006} studied the clustering properties of
galaxies at $z\sim6$ finding an angular correlation for galaxies
brighter than \zz=28.5 mag on scales larger than $10''$ and no
correlation on the same scales going down to \zz=29.0 mag due to low
S/N.  Studying a sample of galaxies at $z\sim5$, \cite{lee2006} found
a further correlation on smaller angular scales, which was interpreted
as due to the so-called one-halo term.  The clustering properties, due
to both the one-halo and the two-halo term, are still not well
explored for \ii-dropouts with magnitudes in the range between \zz=28
mag and \zz=30 mag. Moreover, our field is too small to be affected by
the large angular scale clustering which is better constrained
\citep{overzier2006} than the small scale one. On the basis of this,
we explored the effect of galaxy grouping on the power spectrum
building up a toy model which introduces pairs of sources instead of
randomly distributing them according to a non-zero angular correlation
function only on scales smaller than $10''$.

We assumed the one-halo term of $\omega(\theta)$ found by
\cite{lee2006} for \vv-dropout galaxies to be valid also for
\ii-dropouts and placed the mock galaxies in the simulated frame
accordingly. We found that clustering could amplify the signal from
faint galaxies. In particular, simulations with no clustering leads to
an over-estimation of number of faint galaxies (see Section
\ref{sec:simulations} for more details).  In any case, since the
clustering/grouping properties of \ii-dropouts on the smaller angular
scales are not known at the present epoch, our findings need to be
revisited as soon as a proper angular correlation function is found at
$z\sim6$.

\section{Monte Carlo Simulations}
\label{sec:simulations}
The power spectrum permits us to see the existence of an overall
signal coming from faint galaxies, but we are not able to directly
obtain neither the number of those galaxies nor their magnitudes. To
derive the number and properties of the undetected galaxies
responsible for the signal detected in the FFT we performed a series
of Monte Carlo simulations.

To create the mock {\ii} and {\zz} images of the NICP12 field we
started by deriving the values of the magnitudes, below the detection
limit, and the morphology to assign to the simulated galaxies. Then,
we focused on the way to resemble as well as possible the noise and
the large scale effects characteristic of the data.  Finally, we
compared the power spectra obtained from the ACS data with the results
coming from the simulations. We iterated the simulation 100 times to
estimate the variability of the power spectra depending on the
elements we used.

\subsection{Elements of the Simulations}
In the following all the elements used in the simulations will be
fully discussed.

{\it Random Noise:} We used the IDL RANDOMN routine to create a frame
with the same size of our images and randomly distributed values with
a mean of zero and a standard deviation of one. Subsequently, to
reproduce the typical noise of the ACS data, we multiplied this frame
by the characteristic standard deviation of the {\ii} and {\zz} images
we measured with the IRAF\footnote{Imaging Reduction and Analysis
  Facilities (IRAF) is distributed by the National Optical Astronomy
  Observatories which are operated by the Association of Universities
  for Research in Astronomy (AURA) under cooperative agreement with
  the National Science Foundation.} task IMSTATISTICS. The resulting
two random noise frames have a white power spectrum and reproduce
exactly the trend observed at high wave numbers in Figure \ref{PS_iz}.

\begin{figure}
\centering
\includegraphics[scale=0.45]{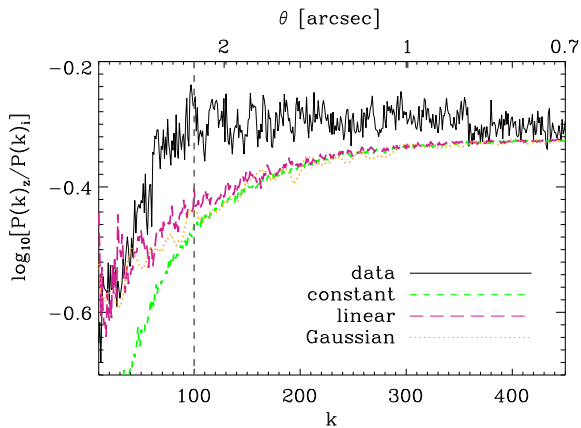}
\caption{Comparison of the ratios between the {\zz} and {\ii} power
  spectra obtained from simulations including only noise and large
  scale residuals. Different models of large scale residuals were
  tested: a constant (green short dashed line), a linear one (magenta
  long dashed line), and a Gaussian one (orange dotted line). In black
  is plotted the ratio between the power spectra derived from the
  data. The dashed vertical line indicates the limit of the wave
  numbers range affected by the type of model considered for the large
  scale residuals. }
\label{comparison_flat}
\end{figure}

{\it Large Scale Effects:} The residuals of the flat field correction
produce the peak at low wave numbers in both the {\ii} and {\zz} power
spectra (Figure \ref{PS_iz}).  Since we were not interested in
studying the trend in the low wave number range, we simply searched a
way to recover the shape of the power spectrum in this range. We
created a model of flat field residuals to reproduce the large scale
effects and, then, we added it to the simulated frame (additive flat
field residuals). To be sure our model was not affecting the results
of the simulations, we performed tests using mock frames with only
noise and large scale residuals, assuming different models for the
latter. In particular, we considered a Gaussian model, a uniform model
(i.e. a constant flat), and one with a linear slope. As can be seen in
Figure \ref{comparison_flat} the differences in the results are
confined in the very low wave number range ($k<100$), up to the dashed
vertical line and it is not possible to reproduce the signal excess
due to faint galaxies with only noise and flat field residuals.  On
the basis of these results we can affirm that the adopted model is not
affecting the range we are interested in, and the signal in
$100<k<400$ is not affected by our choice regarding the modeling of
the residuals of the flat field correction.

\begin{figure}
\centering
\includegraphics[scale=0.45]{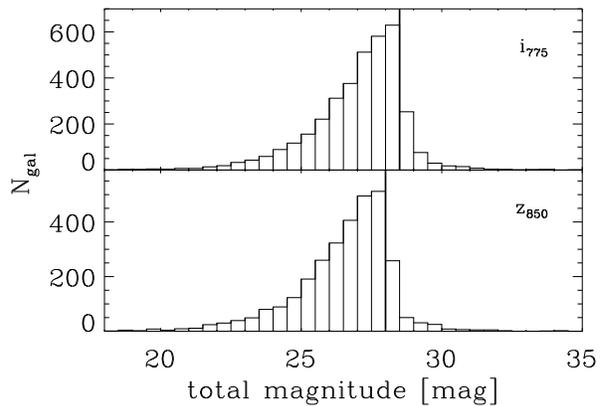}
\caption{Distribution of the total magnitude of the detected galaxies
  in the {\ii} (top panel) and {\zz} (bottom panel) images. The
  vertical line corresponds to the magnitude where we observe a drop
  in the galaxy number and which was assumed as the brightest
  magnitude used in our simulations.}
\label{histo_mag_auto}
\end{figure}

{\it Faint Galaxies:} 
To determine the magnitudes of the simulated faint galaxies, we
considered all the detected galaxies of the field. The histogram of
the distribution of the values of the total magnitude (obtained from
the MAG\_AUTO parameter in SExtractor) for the {\ii} image is shown in
the top panel of Figure \ref{histo_mag_auto}. The value of 28.5 mag,
where the number of galaxies decreases by a factor larger than 50\%,
is assumed as the brighter limit for the range of magnitudes used in
the simulation of the {\ii} image. The fainter limit is $i_{775}=30.5$
mag, as we will see later in this Section.

\begin{table}
\centering
\caption{Number and magnitudes of the simulated galaxies.}
\begin{tabular}{crcr}
\hline
\hline
\multicolumn{2}{c}{$ z\le 5$, $\alpha_i=-1.3$} & \multicolumn{2}{c}{$ z= 6$, $ \alpha=-1.9$}\\
\hline
Mag & Number& Mag & Number\\
\hline
28.5 &  714  & 28.0 & 32\\
29.0 &  1605 & 28.5 & 87\\
29.5 &  3008 & 29.0 & 233\\ 
30.0 &  5503 & 29.5 & 622\\ 
30.5 &  10028& 30.0 & 1665\\ 
\hline
\label{tab_num_mag}
\end{tabular}
\end{table}

The number of faint galaxies with $z \le 5$ to be inserted in the mock
{\ii} image was derived extrapolating the trend of the number counts
beyond the cut off assuming a slope $\alpha_i=-1.3$ according to
\cite{metcalfe1995}. The number of galaxies for each bin of magnitude
used in the simulation is specified in Table \ref{tab_num_mag}.

To reproduce the {\zz} image, we considered both the $z \sim6$ faint
galaxies and those at $z \le 5$. The latter include the same mock
galaxies used to simulate the {\ii} image, whose magnitudes were
corrected with a color term randomly drawn from a distribution
reproducing the {\ii}$-${\zz} color observed for the detected
galaxies. The former population consists of $z \sim6$ galaxies not
detectable in the {\ii} image.  The brighter limit for the magnitude
range is assumed to be the value of 28 mag where the histogram of the
total magnitudes shows a drop larger than 50\% (Figure
\ref{histo_mag_auto}, bottom panel). The fainter limit of the range
considered is $z_{850}=30.5$ mag.  We extrapolated to fainter
magnitudes the luminosity function at $z \sim6$ to obtain the
number of galaxies to be created. 

\begin{figure}
\centering
\includegraphics[scale=0.45]{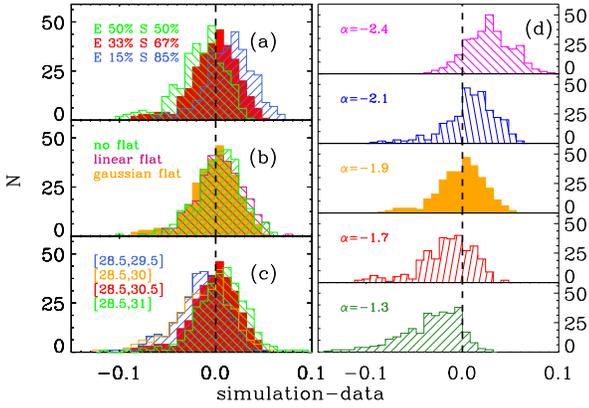}
\caption{Distribution of the residuals derived computing the
  difference between the different realization of the simulation and
  the data, in terms of the ratio between power spectra in the range
  between k=100 and k=400.  The top left panel, (a), refers to the
  different percentages of spirals and ellipticals, the central left one,
  (b), to the various flat field residual models, the bottom left one,
  (c), concerns the magnitude cut-off, and the right one, (d), is
  about the different value of the faint end slope $\alpha$. The
  filled histograms refer to the selected realization, i. e. the one
  with the lowest $\chi^2$.  }
\label{histograms_residuals}
\end{figure}

All the simulations hereafter include a flat field residual model.
After a series of tests, we found a good agreement in reproducing this
effect by combining 20 Gaussian functions with a standard deviation of
about one third of the image dimension: 10 of them have a peak of
$10^{-4}$ to match the typical fluctuations of the flat field, and the
other 10 Gaussians have a peak of $-10^{-4}$ to create a smooth
background with soft large scale fluctuations.  The chi-square
statistics results obtained comparing the data to the simulations
including the three different flat-field models plus mock galaxies and
noise in the range $100< k <400$ were very similar (see Columns 3 and
4 in Table \ref{chi_squared} and panel (b) of Figure
\ref{histograms_residuals}).

\begin{figure}
\centering
\includegraphics[scale=0.45]{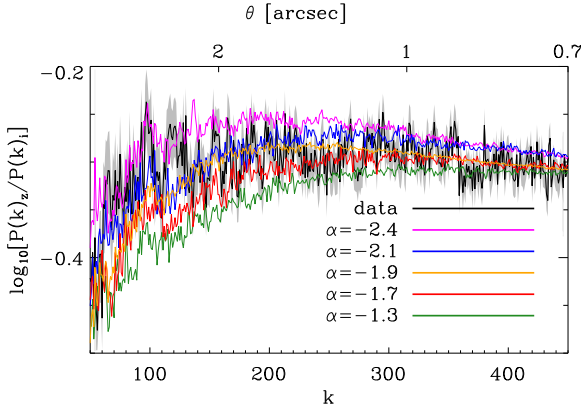}
\caption{Comparison of the ratio between the simulated power spectra
  derived assuming different faint end slopes. The green, red, orange,
  blue, and magenta  lines represent the  median obtained
  from simulations assuming $\alpha=-1.3$, $\alpha=-1.7$,
  $\alpha=-1.9$, $\alpha=-2.1$, and $\alpha=-2.4$, respectively. In
  black the ratio derived from the data is plotted with its error bars.}
\label{comparison_alpha}
\end{figure}

We tested different values for the slope $\alpha$, in particular
$\alpha=-1.3$, $\alpha=-1.7$, $\alpha=-1.9$, $\alpha=-2.1$, and
$\alpha=-2.4$. On the basis of a $\chi^2$ statistics (see columns 7
and 8 of Table \ref{chi_squared} and (b) panel of Figure
\ref{histograms_residuals}) we assumed our extrapolation to have the
slope $\alpha=-1.9$ (Figure \ref{comparison_alpha}) and we anchored it
to the LF value at a magnitude of 28 mag given by
\cite{bouwens2007}. This value is in agreement with the expectation of
\cite{su2011}, who found the faint end slope of the LF at $z \sim6$ to
be in the range $-1.90<\alpha<-1.55 $.

\begin{figure}
\centering
\includegraphics[scale=0.45]{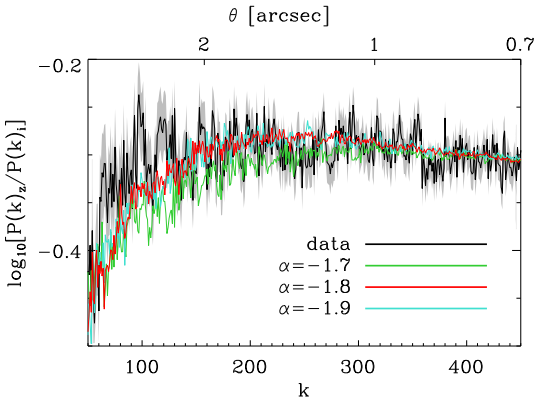}
\caption{Comparison of the ratio between the simulated power spectra
  derived assuming different faint end slopes and pairs of galaxies
  with a distance derived from the angular correlation function at
  $z\sim5$ \citep{lee2006}. The green, red, and blue lines represent
  the median obtained from simulations assuming $\alpha=-1.7$,
  $\alpha=-1.8$, and $\alpha=-1.9$, respectively. In black the ratio
  derived from the data is plotted with its error bars.}
\label{comparison_alpha_clustering}
\end{figure}

Testing the effect of groups of galaxies as described in Section
\ref{Sources of Contamination}, we considered different slopes
$\alpha$, i.e.  $\alpha=-1.7$, $\alpha=-1.8$, and $\alpha=-1.9$
introducing pairs of mock galaxies in the simulation (Figure
\ref{comparison_alpha_clustering}). The $\chi^2$ statistics shows that
$\alpha=-1.8$ fits best the data ($\chi_\nu^2=1.06$, see Columns 9 and
10 of Table \ref{chi_squared}).

Finally we also considered the effect of contaminants at lower
redshift changing the low-luminosity slope of the distribution of
galaxies with $z\le 5$.  To this aim we considered higher and lower
values for $\alpha_i$ and we noted that this change produces a similar
one, with opposite sign, in the derived $\alpha$ value at $z\sim6$.

\begin{figure}
\centering
\includegraphics[scale=0.45]{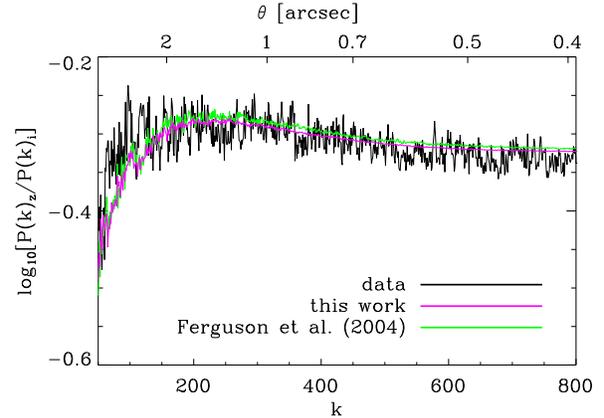}
\caption{Comparison between the ratio of the power spectra obtained
  from the data (black line) and from the simulations assuming our
  effective radius distribution  (magenta
  line) and the one by Ferguson et al. (2004, green line)}
\label{comparison_re}
\end{figure}

In building the set of galaxies, we considered a range of geometrical
(i.e., position and orientation on the frame) and structural
parameters (i.e., effective radius, $r_{\rm e}$, and axis ratio,
$b/a$). Each synthetic galaxy was generated with a set of random,
equally distributed, values for the former parameters and a set of
values which resembles the characteristics of the fainter detected
galaxies of the NICP12 field, derived from the catalog obtained
running SExtractor, for the latter.  Regarding the former parameters,
each synthetic galaxy was generated with a set of random values.  On
the other hand, the structural parameters values had to resemble the
characteristics of the fainter detected galaxies of the NICP12 field,
derived running SExtractor.  The values of the effective radius were
chosen within the range $1.5 \le r_{\rm e} \le 11$ pixel corresponding
to $0\farcs 135 \le r_{\rm e} \le 0\farcs 99$,
and the range spanned by the axis ratio was $0.3 \le b/a \le 1$. Even
though the distribution of $r_{\rm e}$ we used was only slightly
different from the one found by \citet{ferguson2004} at $z\sim5$, we
tested the effect of this parameter on the simulations (Figure
\ref{comparison_re}). In particular, the reduced chi-square statistics
($\chi_\nu^2=1.22$) supports the idea that small differences in the
$r_{\rm e}$ distribution do not significantly affect the result.

Finally, we convolved the mock galaxies with the HUDF PSF
($\rm{FWHM} \sim 0\farcs 1$, \citealt{oesch2007}).

\begin{figure}
\centering
\includegraphics[scale=0.45]{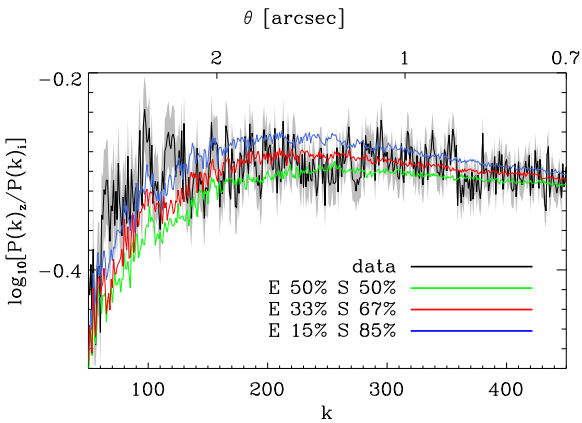}
\caption{Comparison of the ratio between the {\zz} and {\ii} power
  spectra (black line) with the median derived from the simulations
  with a different percentage of spirals and ellipticals represented
  by different colors, 50\% ellipticals and 50\% spirals in green,
  33\% ellipticals and 67\% spirals in red, and 15\% ellipticals and
  85\% spirals in blue. }
\label{ratio_percentage}
\end{figure}

{\it Galaxy Morphology:} Regarding the morphology of the mock
galaxies, we adopted a S\'ersic profile and, changing the S\'ersic
index $n$, we obtained ellipticals ($n=4$) and spirals ($n=1$).
Since there are no clear indications regarding the distribution of
different galactic types in the early Universe, we studied the effect
of changing the relative number of ellipticals and spirals in our
simulations.  We compared the results obtained introducing 15\%
ellipticals (described by a de Vaucouleurs profile) and consequently
85\% spirals (described by an exponential profile) in the simulations
with those obtained simulating 50\% ellipticals and 50\% spirals and
33\% ellipticals and 67\% spirals. In Figure \ref{ratio_percentage}
the ratio of the power spectra obtained from the data is compared with
the mean ratios and $3\sigma$ confidence ranges derived from
simulations with different distributions of ellipticals and
spirals.  Applying the chi-square statistics to the wave number
  range where the peak is, i.e. $100 \lesssim k \lesssim 400$ (see
  Columns 1 and 2 in Table \ref{chi_squared}) we selected the
  realization characterized by the smallest $\chi^2$ value as the best
  fit (see also the distribution of the residuals in the (a) panel of
  Figure \ref{histograms_residuals}). It corresponds to the mix of
  33\% ellipticals and 67\% spirals. These percentages are consistent
with the findings of \cite{somerville2001}, who predicted the majority
of galaxies at $z \geq 3$ to be disk dominated, and of \cite{lotz2006}
and \cite{ravindranath2006} who found the percentage of galaxies with
bulgelike morphology to be $\sim30$\% at $z\sim3$ and $z\sim4$,
respectively.

\begin{table*}
\centering
\caption[$\chi^2$ statistics]{The results obtained from the $\chi^2$
  statistics are listed for all the realizations of the simulations
  regarding the percentage of elliptical and spiral galaxies (Columns
  1 and 2), flat field residual model (Columns 3 and 4) introduced,
  value of the cut-off in magnitude (Columns 5 and 6), and slope of
  the faint end of the LF with single sources (Columns 7 and 8) and
  with pairs of galaxies (Columns 9 and 10), respectively.  The
  selected realization is the one characterized by the smallest
  $\chi_{\nu}^2$ value, i.e. $\chi_{\nu}^2=1.02$. A value of
  $\chi^2_{\nu} =1.24$ corresponds to a probability of 0.3\% and it
  can be rejected with a 3$\sigma$ confidence level. }
\begin{tabular}{cccccccccc}
\hline \hline  
\multicolumn{2}{c}{Percentage } & \multicolumn{2}{c}{Flat field }& \multicolumn{2}{c}{Magnitude }& \multicolumn{2}{c}{Faint end}& \multicolumn{2}{c}{Grouping}\\
\multicolumn{2}{c}{of E and S} & \multicolumn{2}{c}{residual model}& \multicolumn{2}{c}{cut-off}& \multicolumn{2}{c}{slope $\alpha$}& \multicolumn{2}{c}{slope $\alpha$}\\
\hline  
sim & $\chi^2_{\nu}$ & sim & $\chi^2_{\nu}$&sim & $\chi^2_{\nu}$&sim & $\chi^2_{\nu}$&sim & $\chi^2_{\nu}$\\
\hline 
15\%E  & 1.47 & Gaussian& 1.02 & 29.5 mag& 1.60 &$\alpha$=-1.3& 3.22&$\alpha$=-1.7&1.60\\ 
33\%E  & 1.02 & linear  & 1.04 & 30.0 mag& 1.70 &$\alpha$=-1.7& 1.65&$\alpha$=-1.8&1.06\\ 
50\%E  & 1.49 & constant & 1.04 & 30.5 mag& 1.02 &$\alpha$=-1.9& 1.02&$\alpha$=-1.9&1.17\\
&&&                            & 31.0 mag& 1.33 &$\alpha$=-2.1& 1.47&&\\
&&&&&&$\alpha$=-2.4& 2.53\\
\hline
\end{tabular}
\label{chi_squared}
\end{table*}

\begin{figure}
\centering
\includegraphics[scale=0.45]{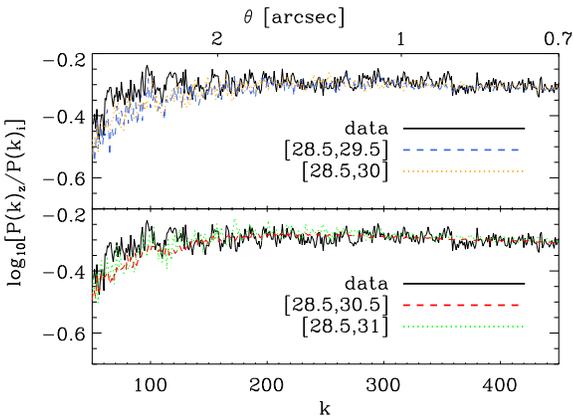}
\caption{Comparison of the observed ratio between the {\zz} and {\ii}
  power spectra (black) and those obtained by simulating unresolved
  galaxies in different magnitude ranges, i.e., $28.5<i_{\rm
    775}<29.5$ (blue dashed line), $28.5<i_{\rm 775}<30$ (orange
  dotted line), $28.5<i_{\rm 775}<30.5$ (red dashed line), and
  $28.5<i_{\rm 775}<31$ (green dotted line). }
\label{multiplot_comparison_magnitudes}
\end{figure}

This mix was adopted to test the sensitivity of the method to the
assumed value of the faint end limit. To this aim we performed a few
simulations changing the fainter magnitude limit of the mock galaxies
(Figure \ref{multiplot_comparison_magnitudes}).  We used the
chi-square statistics to select the best simulation concerning the
magnitude cut-off (see Columns 5 and 6 in Table
\ref{chi_squared}). Moreover, we computed the difference between the
simulations and the data, and we plotted the residuals (Figure
\ref{histograms_residuals}, (c) panel ). On the basis of above, we
decided to create galaxies with magnitudes up to 30.5 mag.

{\it The Mask:} Consistent results can be achieved only by analyzing
the same area in data and simulations. For this reason we used, for
the simulated images, the same mask created from the ACS
data. Moreover, we considered what percentage of the frame is masked
and calculated the number of mock galaxies that had to be outside the
masked area.  We, therefore, placed them randomly on the frame
checking that they were positioned outside the mask.

The sum of the simulated galaxies, the frame reproducing the large
scale effects, and the random noise frames was multiplied by the mask
and analyzed with the power spectrum technique.
\subsection{Results of the Simulations}
\begin{figure}
\centering
\includegraphics[scale=0.45]{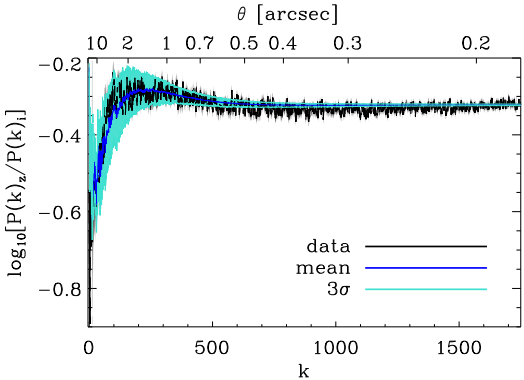}
\caption{Mean ratio between the {\ii} and {\zz} simulated power
  spectra (blue line) and $3\sigma$ confidence range (light blue
  region) derived from 100 simulations compared to the results coming
  from the ACS data (black line).}
\label{ratio_100sim}
\end{figure}

The results obtained by iterating the simulations 100 times considering
the 5 bins of magnitudes given in Table \ref{tab_num_mag} and assuming
one third of the galaxies to be ellipticals and the others spirals are
shown in Figure \ref{ratio_100sim}. The trend of the data with the
associated errors is well fitted by the mean ratio between the
simulated power spectra, and all the values are in agreement with the
$3\sigma$ confidence range.

\subsection{The z $\sim6$ Faint Galaxies Contribution to Cosmic Reionization}
\label{reionization}
 In order to evaluate the implications for reionization of our
 findings, we considered the total surface brightness contribution of
 the undetected galaxies following the approach of
 \cite{stiavelli2004a}. When this contribution is added to that of
 detected galaxies, we find a surface brightness of 24.8 mag
 arcmin$^{-2}$ when extrapolating to a magnitude $z_{\rm 850}=29$ mag
 and 23.6 mag arcmin$^{-2}$ when extrapolating to $z_{\rm 850}=30$
 mag. We compared these values to the minimum surface
 brightness of 27.2 mag arcmin$^{-2}$ needed for reionization by
 Population II stars \citep{stiavelli2004b}. Our results give
 comfortable margins of a factor 9 and 27, respectively, to
 accommodate the escape fraction $f_c$ and the clumping factor
 $C$. For instance, assuming $C=6$ and the Population II spectral
 energy distribution (SED) of \cite{stiavelli2004b}, the LF
 extrapolated down to $z_{\rm 850}=30$ produces enough ionizing
 photons to reionize the Universe if $f_c \gtrsim 5$\%.

So far the implications for reionization of what we found seem to be
very promising, but we are waiting for a more thorough analysis of all
available fields before pushing this interpretation any further.

\section{Conclusions}

In this paper we presented a new technique to quantify the light
contribution coming from the faint high redshift galaxies below the
detection threshold of imaging data.  Specifically, we illustrated the
technique with an application to ACS data obtained in the F775W and
F850LP filters for the HUDF parallel field NICP12 looking for faint
$z\sim6$ galaxies on the basis of the Lyman break technique, which is
based on a comparison between the flux of a galaxy within the set of
two filters.

In particular, our aim was to extend by a few magnitudes the faint end
of the luminosity function at that redshift. After masking all the
detected sources in the field we applied a Fast Fourier Transform to
obtain the spatial power spectrum of the background signal and we
examined the ratio between the {\zz} and {\ii} power spectra. In this
way we were able to distinguish the sources of noise from the
fluctuations due to the unresolved galaxies. We tested the effect of
several parameters on the power spectrum to test its reliability and
how contamination can affect the results.  

We showed that the power spectrum technique is a powerful tool in
analyzing the light contribution produced by galaxies which are below
the detection limit in deep and ultra deep surveys. We used this
technique to estimate the contribution to cosmic reionization from
faint galaxies ($z_{\rm 850} \ge 28$ mag), which are bona fide $z\sim
6$ candidates, in the NICP12 ACS field of the UDF05.

Monte Carlo simulations were used to determine the number of faint
undetected $i$-dropouts responsible for the peak observed in the ratio
between the {\zz} and {\ii} power spectra (Figure \ref{PS_ratio}). The
data are consistent with a faint end slope of the luminosity function
of $\alpha = -1.9$ if not introducing any clustering/grouping in the
\ii-dropouts distribution, and $\alpha = -1.8$ when considering pairs
of galaxies with distances drawn accordingly to the angular
correlation function by \cite{lee2006}.  Both the $\alpha$ values
  we obtained are consistent with previous findings by
  \cite{yanwindhorst2004}, \cite{malhotra2005}, \cite{bouwens2006},
  \cite{bouwens2007}, \cite{bouwens2011b}, and \cite{su2011}.

Considering $\alpha = -1.9$, adding the total surface brightness
contribution of faint galaxies to that from the detected galaxies and
comparing it to the minimum value required to reionize the Universe,
we obtained a margin to model the escape fraction and the clumping
factor.  Adopting the clumping factor of 6 and Population II SED of
\cite{stiavelli2004b}, the $z\sim6$ undetected galaxies down to
$z_{\rm 850}=30$ mag produce enough ionizing photons to reionize the
Universe assuming the escape fraction to be larger than $\sim 5$\% The
solution with all galaxies in pairs and a slope $\alpha = -1.8$
changes the cumulative surface brightness by less than 0.1 mag and
implies an escape fraction larger by $\sim10\%$. Thus, the main
conclusion of the paper that reionization is indeed consistent with
being completed due to the contribution of faint galaxies is not
affected by the presence of multiple halo occupation.

We expect to apply our technique to the ultra deep data obtained with
NICMOS and the Wide Field Camera 3 (WFC3). The comparison of the
results for different fields and instruments will permit us to better
constrain the role that faint galaxies at $z \sim6$, $ z \sim7$, and
$z \sim8$ had in the cosmic reionization.

\section*{Acknowledgments}
We acknowledge useful conversations with Mattia Negrello and Nelson
Padilla.  We thank the anonymous referee for his/her comments that
helped to improve our paper.  This paper was supported by Padua
University through grants CPDA089220/08 and CPDR095001/09, by Italian
Space Agency through grant ASI-INAF I/009/10/0, by HST go grant O1374,
and the JWST ids grant NAG5-12458.  VC is grateful to Space Telescope
Science Institute for its hospitality while part of this paper was
being written. LM is supported by the University of Padua through
grant No. CPS0204.

\label{lastpage}

\end{document}